\documentclass[letterpaper, english, aps, pra, twocolumn, floatfix, showpacs, superscriptaddress, longbibliography]{revtex4-1}
\usepackage{epsfig, graphicx, float}
\usepackage[T1]{fontenc}
\usepackage[latin1]{inputenc}
\usepackage{bm}
\usepackage{babel}
\usepackage{hyperref}
\usepackage{amsmath}
\usepackage{amssymb}
\usepackage{color}
\begin{document}

\title{Density-dependent hopping for ultracold atoms immersed in a Bose-Einstein-condensate vortex lattice}
\author{R. H. Chaviguri}
\affiliation{Instituto de F\'{i}sica de S\~ao Carlos, Universidade de S\~ao Paulo, C.P. 369, S\~ao Carlos, SP, 13560-970, Brazil}
\author{T. Comparin}
\affiliation{INO-CNR BEC Center and Dipartimento di Fisica, Universit\`{a} di Trento, 38123 Povo, Italy}
\author{M. Di Liberto}
\affiliation{INO-CNR BEC Center and Dipartimento di Fisica, Universit\`{a} di Trento, 38123 Povo, Italy}
\affiliation{Center for Nonlinear Phenomena and Complex Systems, Universit\'{e} Libre de Bruxelles, CP 231, Campus Plaine, B-1050 Brussels, Belgium}
\author{M. A. Caracanhas}
\affiliation{Instituto de F\'{i}sica de S\~ao Carlos, Universidade de S\~ao Paulo, C.P. 369, S\~ao Carlos, SP, 13560-970, Brazil}

\begin{abstract}
Both mixtures of atomic Bose-Einstein condensates and systems with atoms trapped
in optical lattices have been intensely explored theoretically, mainly due to
the exceptional developments on the experimental side.  We investigate the
properties of ultracold atomic impurities (bosons) immersed in a vortex lattice
of a second Bose-condensed species. In contrast to the static optical-lattice
configuration, the vortex lattice presents intrinsic dynamics given by its
Tkachenko modes. These excitations induce additional correlations between the
impurities, which consist of a long-range attractive potential and  a
density-dependent hopping, described here in the framework of an extended
Bose-Hubbard model.  We compute the quantum phase diagram of the impurity
species through a Gutzwiller ansatz and through the mean-field approach, and
separately identify the effects of the two additional terms, i.e., the shift and
the deformation of the Mott insulator lobes.  The long-range attraction, in
particular, induces the existence of a triple point in the phase diagram, in
agreement with previous quantum Monte Carlo calculations [Chaviguri \emph{et
al.}, Phys. Rev. A \textbf{95}, 053639 (2017)].
\end{abstract}

\date{\today}
\maketitle

\section{Introduction}

Ultracold atoms in optical lattices provide access to a rich set of interesting
quantum-physics systems. This is especially due to the high experimental control
achievable and to the possibility of reaching strongly-correlated phases
~\cite{Bloch2005NP, Bloch2008RMP, Lewenstein2012}.  Both bosonic and fermionic
atomic species can be trapped in optical lattices, with several allowed
geometries and the possibility of engineering different kinds of interactions.
The paradigmatic example is the case of spinless bosons with short-range
interactions, described via the ordinary Bose-Hubbard (BH) model
~\cite{Fisher1989PRB, Jaksch1998PRL, VanOosten2001PRA}. Within this model,
hopping between sites favors a delocalized phase, while the on-site interatomic
repulsion suppresses density fluctuations and favors localization.  Such
competition results in a quantum phase transition between the Mott-insulator
(MI) and superfluid (SF) phases, and its signature was clearly identified in a
breakthrough experiment~\cite{Greiner2002}.  Among the several possible
extensions beyond this example, we focus on the addition of long-range
interactions and of density-dependent hopping.

In the original BH model, only interactions between atoms on the same site are
considered, as the interatomic potential is already negligible at the distance
of one lattice constant.  For longer-ranged interactions, however, one also has
to include atomic pairs at larger distance, starting with those on neighboring
sites.  This is the case for atomic species with large dipolar moment, for which
the long-ranged dipole-dipole interaction is present.  The corresponding
extended Bose-Hubbard model (EBH) was recently realized in experiments with a
gas of erbium atoms in a three-dimensional lattice~\cite{Baier2016}.  On the
theoretical side, there exist several predictions for systems with long-range
interactions.  In one dimension, they include the peculiar Haldane-insulator
phase~\cite{DallaTorre2006PRL}, while in higher dimensions they range from
density waves to supersolidity, for both hard-core and soft bosons -- see
reviews in Refs.~\cite{Lahaye2009RPP, Trefzger2011JPB}.

A different extension of the BH model consists of adding terms where the hopping
between two sites also depends on the two corresponding densities.  In the study
of the Fermi-Hubbard model, this density-dependent hopping was introduced to
study ferromagnetism and superconductivity in solid-state materials
~\cite{Amadon1996PRB, Hirsch1989PC, Hirsch1994PB}.  For ultracold atoms in
optical lattices, it is a term that is often negligible but in principle always
present, as it is related to off-site matrix elements of the interaction term.
In some cases, including the case of strong dipolar interactions, this term is
non-negligible, as predicted by theory~\cite{Jurgensen2012PRA, Luhmann2012NJP}
and observed experimentally~\cite{Jurgensen2014PRL, Baier2016}.  Alternatively,
one can artificially enhance the density-dependent hopping term via Floquet
driving schemes.  One possibility is given by a time-dependent modulation of the
$s$-wave scattering length~\cite{Rapp2012PRL, DiLiberto2014PRA}, later
experimentally realized ~\cite{Meinert2016PRL}. Another possibility is given by
near-resonant lattice shaking~\cite{Gorg2017Arxiv}.

In this and previous works~\cite{Johnson2016PRL, Chaviguri2017PRA}, the trapping
mechanism provided by an optical lattice is replaced by the underlying vortex
lattice generated in a Bose-Einstein condensate (BEC). The experimental
production of vortices in ultracold dilute gases varies from a few
units~\cite{Madison2000PRL, Haljan2001PRL} to large arrays~\cite{AboShaeer2001},
arranged in the Abrikosov triangular configuration.  The peculiar dynamics of
the vortex lattice is characterized by the Tkachenko vibrational
modes~\cite{Tkachenko1969JETP}, which were also identified in ultracold-atoms
experiments~\cite{Coddington2003PRL}. Such a lattice can trap atomic impurities
(that is, atoms of a different species), which are then described through a
discrete lattice model

Taking into account the dynamics of the vortex lattice, in this work we extend
the effective model for the atomic impurities of Ref.~\cite{Chaviguri2017PRA}.
The derivation of this extended Bose-Hubbard Hamiltonian (EBH) is based on the
polaron transformation, with parameters chosen through a variational approach,
and the resulting model includes both the nontrivial ingredients mentioned
above: A long-range interaction and a density-dependent hopping.

To explore the new features of this system we determine its phase diagram
through approximate methods (based on the Gutzwiller \emph{ansatz} or on a
mean-field decoupling), and identify the effects of the two additional terms in
the Hamiltonian.  The long-range attraction is known to induce a change in the
position and size of the MI regions in the phase diagram.  By extending the
analysis of Ref.~\cite{Chaviguri2017PRA}, we observe that it also introduces a
discontinuous transition between Mott insulators with different filling, ending
at a triple point where it merges with the conventional MI/SF phase boundary.
The density-dependent hopping has a positive coefficient so that its main effect
is to enhance superfluidity. At a difference with the nearest-neighbor
attractive interaction, this effect results in a shift of the critical hopping
parameter for the MI/SF transition.

The paper is structured as follows: Section~\ref{section:physicalmodel}
corresponds to the derivation of an effective model for atomic impurities
immersed in the a vortex lattice BEC, while the resulting EBH model is
characterized in Section~\ref{section:EBH}, followed by the conclusions in
Section~\ref{sect:conclusions}.

\section{Physical model}
\label{section:physicalmodel}

We consider a quasi-2D system composed of two ultracold atomic species, the
majority species $A$ and the impurity species $B$.
The Hamiltonian of the system is the sum of three terms~\cite{Chaviguri2017PRA},
\begin{eqnarray}
\label{1}
\nonumber
{H}_A &=&
\int d^2r\,\Big[
\hat\psi_A^\dagger(\mathbf{r})
\frac{(-i\hbar\nabla-{\bf A(\mathbf{r})})^2}{2 m_A}
\hat\psi_A^{\phantom\dagger}(\mathbf{r}) + \\
\nonumber
&&
+ \hat\psi_A^\dagger(\mathbf{r}) V_{\textrm{ext}}(\mathbf{r}) \hat\psi_A(\mathbf{r})
+ \frac{g_{A}}{2} \left(\hat\psi_A^{\dagger}(\mathbf{r})\hat\psi_A(\mathbf{r})\right)^2 \Big], \\
\nonumber
{H}_B &=&
\int d^2r\,\Big[
\hat\psi_B^\dagger(\mathbf{r})
\frac{(-i\hbar\nabla)^2}{2 m_B}
\hat\psi_B(\mathbf{r}) + \\
\nonumber
&&
+ \hat\psi_B^\dagger(\mathbf{r}) V_{\textrm{ext}}(\mathbf{r}) \hat\psi_B(\mathbf{r})
+ \frac{g_B}{2} \left(\hat\psi_B^{\dagger}(\mathbf{r}) \hat\psi_B(\mathbf{r})\right)^2 \Big], \\
{H}_{AB} &=& g_{AB} \int d^2r\,
\hat\psi_A^{\dagger}(\mathbf{r}) \hat\psi_B^{\dagger}(\mathbf{r})
\hat\psi_A (\mathbf{r}) \hat\psi_B(\mathbf{r}),
\end{eqnarray}
where $\hat{\psi}^{\dagger}_{i}(\mathbf{r}) \; (\hat{\psi}_{i}(\mathbf{r}))$ is
the creation (annihilation) operator, and $m_{A}$ and $m_{B}$ are the atomic
masses of the two species.  The two-dimensional intra- and inter-species contact
interactions, $g_{i} = {2\sqrt{2\pi} \hbar^2\,a_{i}}/{m_{i} l^i_{z}}$ and
$g_{AB} = {\sqrt{2\pi}\hbar^2\,a_{AB}}/{m_{AB} l^{AB}_{z}}$, depend on the
corresponding scattering lengths $a_i$ and $a_{AB}$ (for $i\in\lbrace
A,B\rbrace$), and the reduced mass reads $m_{AB}={m_A m_B} / {(m_A + m_B)}$.
The transverse harmonic confinement defines the characteristic lengths
$l^{i}_{z} = \sqrt{\hbar / (m_{i}\omega_{z})}$ and $l^{AB}_{z} =
\sqrt{\hbar / (2m_{AB}\omega_{z})}$.

In addition, a synthetic magnetic field is introduced for $A$ atoms. Within the
scheme described in Ref.~\cite{Spielman2009PRA}, laser beams are used to couple
internal atomic states, and a magnetic-field gradient provides the inhomogeneity
required to generate a non-trivial synthetic field. In the Hamiltonian, this
field is represented by the pseudovector potential $\mathbf{A}$, which can be
written as $\mathbf{A} = m_{A}\mathbf{\Omega} \times \mathbf{r}$ (with
$\mathbf{\Omega}$ pointing in the direction orthogonal to the plane).  The
technique in Ref.~\cite{Spielman2009PRA} would also modify the confining
potential $V_\mathrm{ext}$, but we neglect this effect here.  Up to a dozen
vortices were generated in the original experiment based on this
technique~\cite{Lin2009}, and a method to increase this number was recently
proposed~\cite{Price2016NJP}. Moreover, numerical solutions of the
Gross-Pitaevskii equation for this system show that an extended vortex lattice
can form~\cite{Spielman2009PRA, Bai2017Arxiv}.

The Gross-Pitaevskii equation can describe a two-dimensional vortex lattice at
$T=0$ in the quantum Hall (QH) regime~\cite{Mueller2002PRL, Matveenko2011PRA,
Cooper2008AP}.  In this regime, the occupied states are the quasi-degenerate
lowest Landau level $\varphi_{A} \propto \prod_{k}(z-\zeta_{k})$, where
$\zeta_{k} = (x_k+iy_k)/l$ is a complex number which represents the position of
the $k$th vortex in the lattice [with $(x,y) = \mathbf{r}$], here normalized by
the magnetic length $l = \sqrt{\hbar/m_{A}\Omega}$. The wave function of species
$A$ is given by $\psi_{A}(\mathbf{r})=\sqrt{n_A} \varphi_A(\mathbf{r})$, where
$n_A = N_A /S$ is the average atomic density and $S$ is the total surface of the
lattice.

In the QH regime, the vortex lattice is established with the population of
species $A$ much higher than the number of vortices. We require the number of
$B$ impurities to be of the same order as the number of vortices $N_V$, $N_B
\sim N_V \ll N_A$, such that the vortex-lattice structure is not affected by
their presence.

In the grand-canonical formalism, we decompose the field operator of species $A$
into a condensed part and its fluctuations, namely $\hat{\psi}_{A} = \psi_{A} +
\delta\hat{\psi}_{A}$. Substituting in the total Hamiltonian of Eq.~\eqref{1},
and keeping terms up to quadratic order in the fluctuations, yields
\begin{equation}
\label{2}
K =
\underbrace{H_{B}+H_{AB}^{(0)}-\mu_{B}\hat{N}_{B}}_{K_{B}} +
\underbrace{H_{AB}^{(1)}}_{H_\mathrm{int}} +
\underbrace{H_{A}^{(0)}+H_{A}^{(2)}}_{K_A}
\end{equation}
where $\hat{N}_{B}$ is the impurity number operator. The upper index in the
Hamiltonians of Eq.~\eqref{2} indicates the expansion order in terms of the
species $A$ field fluctuation $\delta\hat{\psi}_{A}$. As a reminder, the
validity of the Gross-Pitaevskii equation implies $H_{A}^{(1)} = 0$.

In the following sections, we consider the two cases where the vortex-lattice
fluctuations are neglected or included, namely the static and dynamical
lattice.

\subsection{Static lattice}

Disregarding the vortex lattice fluctuations, i.e., assuming the mean-field (MF)
regime $\hat{\psi}_{A}\approx\psi_{A}$, the contribution of species $A$
Hamiltonian is reduced to a constant energy shift $H_{A}^{(0)}$.  Inter-species
interactions contribute with an effective MF potential given by $H_{AB}^{(0)} =
\int d^2 r\, V_A(\mathbf r) \hat{\psi}_B^\dagger (\mathbf r)
\hat{\psi}_B^{\phantom\dagger}(\mathbf r)$, where $V_A(\mathbf r) = n_A g_{AB}
|\varphi_A(\mathbf r)|^2$.  From the result above, we note that the species $A$
behaves like a ``static'' lattice for the impurity species $B$.

Here we consider the tight-binding regime for the impurities trapped in the
vortex lattice sites, with the lattice potential amplitude $V_0 = n_A g_{AB}$
being much larger than the recoil energy $E_R = \hbar^2 / (2 m_B \xi_A^2)$
~\cite{Caracanhas2013PRL}, where $\xi_A = \hbar / \sqrt{2 m_A n_A g_A}$ is the
healing length of species $A$.  In the quantum Hall regime, the intervortex
distance $d = 2 l$ is approximately equal to  $2 \xi_A$~\cite{Fetter2009RMP}.
In general $d^{2} = 16 \Gamma_{L} \xi_A^{2}$, with $\Gamma_{L} = n_A
g_{A}/2\hbar\Omega (\leq1 $\cite{Schweikhard2004PRL})  being a parameter
associated with the lowest-Landau-level constraint, which is related to the
vortex areal density by $N_V \sim 1 / \pi d^{2}$.

In the tight-binding regime ($V_{0} \gg E_{R}$), the large gap between the first
and second Bloch bands allows us to only consider the lowest band to describe
the behavior of the impurities.  The Bloch wave function $\Phi_{\mathbf{k}}
(\mathbf{r})$ for the impurities in the lattice can be related to the Wannier
function $\omega_{B}(\mathbf{r}_i)$, which is strongly localized on the vortex sites
$\mathbf{R}_{i}$, as $\omega_{B}(\mathbf{r}_i) = (1 / \sqrt{N_V}) \sum_{\mathbf{k}}
\Phi_{\mathbf{k}} (\mathbf{r}) e^{i\mathbf{k} \cdot \mathbf{R}_i}$, where $\mathbf{r}_i
\equiv\mathbf{r} - \mathbf{R}_{i}$ and $\int d^{2}r |\omega_{B}(\mathbf{r}_i)|^2 = 1$.
The field operator can be expanded in the Wannier basis as
\begin{equation}
\label{3}
\hat{\psi}_{B}(\mathbf{r}) =
\sum_i
\omega_{B}(\mathbf{r}_i)\,\hat{b}_{i},
\end{equation}
where $\hat{b}_{i}$ ($\hat{b}^{\dag}_{i}$) destroys (creates) impurity atoms on
site $i$.
After inserting Eq.~\eqref{3} in $K_B$, and considering only on-site interactions
and nearest-neighbor hopping terms, one obtains the Bose-Hubbard Hamiltonian
~\cite{Jaksch1998PRL}
\begin{equation}
\label{4}
K^\mathrm{BH}_B =- J \sum_{\langle i,j \rangle} \hat{b}_{i}^{\dag} \hat{b}_{j} +
\frac{U}{2} \sum_{i} \hat{n}_i(\hat{n}_i-1)-\mu_{B} \sum_{i} \hat{n}_i,
\end{equation}
where $\hat{b}_{i}^{\dag} \hat{b}_{i} = \hat{n}_{i}$ is the local-density
operator and the first sum runs over all nearest-neighbor pairs $(i,j)$.
The hopping amplitude $J$ and the on-site interaction energy $U$ read
\begin{align}
J &= -\int d^{2}r\,
\omega^{\ast}_B(\mathbf{r}_i)
\left[
-\frac{\hbar^{2}\nabla^{2}}{2m_{B}} + g_{AB} n_{A} |\varphi_A(\mathbf{r})|^{2}
\right]
\omega_{B}(\mathbf{r}_j),
\label{5}\\
U &= g_{B} \int d^{2}r \, |\omega_B(\mathbf{r})|^{4}.
\label{6}
\end{align}

\subsection{Dynamical lattice}

The vortex lattice presents normal vibrational modes, the Tkachenko modes, which
can be included in our previous model as quantum fluctuations, i.e., using a
beyond MF approach for species $A$, with $\delta\hat{\psi}_{A}\neq0$. The
presence of fluctuations radically modifies the dynamics of impurities. In
addition to changing the structure of the hopping and on-site energy, their
inclusion generates new off-site terms in the BH Hamiltonian that significantly
affect the quantum phase-diagram, as described in section~\ref{section:EBH}.

We apply the Bogoliubov canonical transformation to the fluctuation field
\begin{equation}
\label{7}
\delta\hat\psi_A(\mathbf{r}) =
\frac{1}{\sqrt{S}}
\sum_{\mathbf{q}} \left[
u_{\mathbf{q}}(\mathbf{r})\hat{a}_{\mathbf{q}} -
v_{\mathbf{q}}(\mathbf{r})\hat{a}_{\mathbf{q}}^{\dagger}
\right],
\end{equation}
with $u_{\mathbf{q}}$ and $v_{\mathbf{q}}$ being specific functions associated
with the vortex lattice and $\hat{a}_{\mathbf{q}}^{\dagger}
(\hat{a}_{\mathbf{q}})$ the creation (destruction) operators of a Tkachenko
mode with momentum $\mathbf{q}$ and energy dispersion $\epsilon_{\mathbf{q}}$
~\cite{Matveenko2011PRA}.
Using the Wannier-function expansion of Eq.~\eqref{3} together with the
Bogoliubov transformation in the respective Hamiltonians of Eq.~\eqref{2}, we
obtain
\begin{equation}
\begin{aligned}
K^\mathrm{EBH}_{B} =& \sum_{\mathbf{q}} \epsilon_{\mathbf{q}}
\hat{a}_{\mathbf{q}}^{\dagger}\hat{a}_{\mathbf{q}}
- J\sum_{\langle i,j \rangle}\hat{b}_{i}^{\dag}\hat{b}_{j}
+ \frac{U}{2}  \sum_{i} \hat{n}_i(\hat{n}_i-1)\\
 -&\mu_{B} \sum_{i}\hat{n}_i
+g_{AB} \sqrt{\frac{n_A}{S}} \sum_{\mathbf{q},ij}
\left[
\Omega^{ij}_{\mathbf{q}} \hat{a}_{\mathbf{q}} +
\bar{\Omega}^{ij}_{\mathbf{q}} \hat{a}^{\dagger}_{\mathbf{q}}
\right]
\hat{b}^{\dag}_i\hat{b}_j,
\end{aligned}
\label{8}
\end{equation}
where
\begin{equation}
\begin{aligned}
\Omega^{ij}_{\mathbf{q}} = \int d^2r
\left[ \varphi_{A} ^{\ast}(\mathbf{r})u_{\mathbf{q}}(\mathbf{r}) -
\varphi_{A}(\mathbf{r})v^{\ast}_{\mathbf{q}}(\mathbf{r})\right]
\omega_{B}^{\ast}(\mathbf{r}_{i})\omega_{B}(\mathbf{r}_{j}),
\\
\bar{\Omega}^{ij}_{\mathbf{q}} = \int d^2r
\left[\varphi_{A}(\mathbf{r}) u^{\ast}_{\mathbf{q}}(\mathbf{r}) -
\varphi_{A}^{\ast}(\mathbf{r}) v_{\mathbf{q}}(\mathbf{r})\right]
\omega_{B}^{\ast}(\mathbf{r}_{i})\omega_{B}(\mathbf{r}_{j}).
\end{aligned}
\label{9}
\end{equation}

To cancel the last term in Eq.\eqref{8}, we apply a unitary transformation
~\cite{Lang1963JETP} that renormalizes the coefficients of the total Hamiltonian
to account for the vibrational modes. This gives a polaronic extended
Bose-Hubbard Hamiltonian for the impurity atoms~\cite{Agarwal2013PRB,
Benjamin2014PRA}.
We consider $\tilde{K}_{B} = e^{-\mathcal{U}} K^\mathrm{EBH}_{B} e^{\mathcal{U}} =
K^\mathrm{EBH}_{B}+[\mathcal{U},K^\mathrm{EBH}_{B}]+
(1/2!)[\mathcal{U},[\mathcal{U},K^\mathrm{EBH}_{B}]]+\dots,$
with
\begin{equation*}
\mathcal{U} =
\frac{1}{\sqrt{S}}
\sum_{\mathbf{q},i} \,\left[\;
e^{i\mathbf{q}\cdot\mathbf{R}_{i}}\alpha_{\mathbf{q},i}
^{*}\;\hat{a}^{\dagger}_{\mathbf{q}} -
\;e^{-i\mathbf{q}\cdot\mathbf{R}_{i}}\alpha_{\mathbf{q},i}\;
\hat{a}_{\mathbf{q}} \right] \hat{n}_j
,
\end{equation*}
where the coefficients $\alpha$ of the transformation will be defined by using a
variational method.  Impurity and lattice-mode operators transform as
\begin{align}
e^{-\mathcal{U}} \hat{b}_j e^{\mathcal{U}} &=
\hat{b}_j
e^{
\frac{1}{\sqrt{S}}
\sum_{\mathbf{q}}\left(
e^{-i\mathbf{q}\cdot\mathbf{R}_j} \alpha_{\mathbf{q},j}\;\hat{a}_{\mathbf{q}} -
e^{i\mathbf{q}\cdot\mathbf{R}_j} \alpha^{\ast}_{\mathbf{q},j}\;
\hat{a}_{\mathbf{q}}^{\dagger}\right)
},\nonumber
\\
e^{-\mathcal{U}} \hat{a}_{\mathbf{q}} e^{\mathcal{U}} &=
\hat{a}_{\mathbf{q}} -
\frac{1}{\sqrt{S}}
\sum_j e^{i \mathbf{q}\cdot\mathbf{R}_{j}} \;
\alpha^{\ast}_{\mathbf{q},j} \; \hat{n}_{j}.
\label{10}
\end{align}
By plugging Eq.~\eqref{10} into Eq.~\eqref{8}, and by considering
$\alpha_{\mathbf{q},j} = ({g_{AB}\sqrt{n_A}}/{\epsilon_{q}})
\Omega_\mathbf{q}^{jj} \exp\left({i\mathbf{q}\cdot\mathbf{R}_j}\right)$, which
was determined through the minimization of the total energy of the system (see
Appendix~\ref{App:variationalmethod}), we obtain
\begin{align}
\tilde{K}_{B}& =
-\tilde{J}\sum_{\langle i,j\rangle}\hat{b}^{\dag}_{i} \hat{b}_{j}+
\frac{\tilde{U}}{2}\sum_{i}\hat{n}_{i}(\hat{n}_{i}-1)-
\tilde{\mu}\sum_{i}\hat{n}_{i}
\nonumber\\
&\quad
-\frac{V}{2}\sum_{\langle i,j\rangle}\hat{n}_{i}\hat{n}_{j}-
P\sum_{\langle i,j\rangle}(\hat{n}_{i}+
\hat{n}_{j})\hat{b}^{\dag}_{i}\hat{b}_{j}.
\label{11}
\end{align}
The coefficients read (see Appendix~\ref{App:EBHmodel})
\begin{align}
\tilde{J} &= Jf^{0}, \label{12}\\
\tilde{U} &= U-2\frac{g^{2}_{AB}n_{A}}{S}
\sum_{\textbf{q}}
\frac{|\Omega^{ii}_{\textbf{q}}|^{2}}{\epsilon_{\textbf{q}}}\label{13}, \\
\tilde{\mu} &=\mu_{B}+\frac{g^{2}_{AB}n_{A}}{S}
\sum_{\textbf{q}}
\frac{|\Omega^{ii}_{\textbf{q}}|^{2}}{\epsilon_{\textbf{q}}}\label{14},\\
V&= 2\frac{g^{2}_{AB}n_{A}}{S}
\sum_{\textbf{q}}
\frac{|\Omega^{ii}_{\textbf{q}}|^{2}e^{-i\textbf{q}\cdot\textbf{d}}}{\epsilon_{\textbf{q}}}\label{15},\\
P&=\frac{g^{2}_{AB}n_{A}}{S}
\sum_{\mathbf{q}}
\frac{1}{\epsilon_{\mathbf{q}}}\bigg(
\Omega^{ij}_{\mathbf{q}}\Omega^{ii\ast}_{\mathbf{q}}+
\overline{\Omega}^{ij}_{\mathbf{q}}\Omega^{ii}_{\mathbf{q}}\bigg)f^{0},
\label{16}
\end{align}
where
\begin{equation}
f^{0} =
\exp\left\lbrace
-\frac{g^{2}_{AB}n_{A}}{2S} \sum_{\textbf{q}}
\frac {\left| \Omega^{ii}_{\textbf{q}} \right|^2}
{\epsilon^{2}_{\textbf{q}}}
\left|1 - e^{-i\textbf{q} \cdot \textbf{d}}\right|^{2}\right\rbrace
.
\end{equation}
Vortex-lattice fluctuations induce the appearance of additional terms in the
effective impurity Hamiltonian [cf. difference between Eqs.~\eqref{4} and
\eqref{11}].  An estimate of $V$ and $P$ for typical values of the physical
parameters is reported in Appendix~\ref{App:continuumlimit}.  The long-range
interaction between impurities located at different sites occurs by means of the
Tkachenko-mode scattering. One can show that this interaction decays rapidly
with the inter-vortex distance, so that we can restrict the potential range to
pairs of nearest-neighboring sites.  Moreover, the Hamiltonian includes a
density-dependent hopping, with amplitude $P$, which was not considered in
Ref.~\cite{Chaviguri2017PRA}.  Also in this case, we shall neglect processes
between sites at distances larger than one lattice constant.

\begin{figure*}[tb]
\centering
\includegraphics[width=\linewidth]{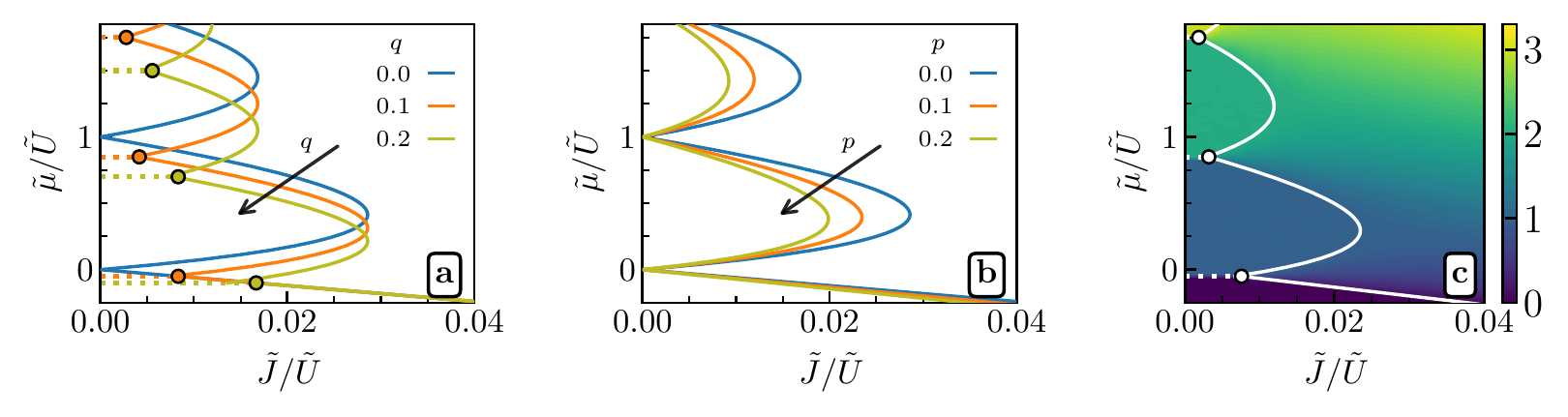}
\caption{
Phase diagram of the EBH model in Eq.~\eqref{11}, featuring continuous MI/SF
transitions (solid lines) and discontinuous MI/MI transitions (dashed lines).
We show the separate effects of the long-range attraction $q$ [panel (a), with
$p=0$] and of the density-dependent hopping $p$ [panel (b), with $q=0$], where
the triple point and the MI/SF boundaries are computed through Eqs.~\eqref{23}
and \eqref{eq:lobes_MF_T0}, respectively.  (c) Density $\langle \hat{n}_i
\rangle$ for $q=p=0.1$, computed through the Gutzwiller method (color codes,
with $n_\mathrm{max} = 6$), compared with the analytical phase boundaries.
\label{fig:phasediags}}
\end{figure*}

\section{Extended Bose-Hubbard phase diagram}
\label{section:EBH}

In this section we determine the ground-state phase diagram of the EBH model in
Eq.~\eqref{11} through two methods.  The Gutzwiller variational approach allows
us to numerically compute relevant observables in the whole phase diagram, to
identify the Mott and superfluid regions. By truncating the number of
variational parameters, we also extract analytical expressions for the
triple-point position and MI/SF phase boundary.  Furthermore, we compute the
MI/SF boundary through a mean-field approach, and the two methods agree with
each other to high accuracy.

We introduce the dimensionless parameters
\begin{equation}\label{19}
q = \frac{zV}{\tilde{U}}, \qquad\qquad
p = \frac{P}{\tilde{J}},
\end{equation}
and define $z$ as the number of nearest neighbors of each site ($z = 6$, for the
triangular lattice). Our results are shown in Figure~\ref{fig:phasediags}, where
we observe several features:
(i)
In the $p = 0$  phase diagram [cf. Fig.~\ref{fig:phasediags}(a)], the
position and size of the MI lobes is modified due to $q>0$, and their boundaries
at $\tilde{J} = 0$ are given by
\begin{equation}
\label{20}
\frac{\tilde{\mu}^{g,g+1}}{\tilde{U}} = g - \frac{2g+1}{2} q,
\end{equation}
for MI lobes with filling $g$ and $g + 1$.
This effect was also described in Ref.~\cite{Chaviguri2017PRA}, and the
boundary between MI lobes at $\tilde{J} = 0$ -- Eq.~\eqref{20} -- is reproduced
both with approximate methods (in the current work) and in the exact quantum
Monte Carlo phase diagram \cite{Chaviguri2017PRA}.
(ii)
The boundary between two subsequent MI regions is not restricted to one point at
$\tilde{J}=0$, but it extends up to a triple point at $\tilde{J} > 0$ [cf. 
Fig.~\ref{fig:phasediags}(a)].
(iii)
A non-zero value of $p$ reduces the area of the MI regions [cf. 
Fig.~\ref{fig:phasediags}(b)], which follows from the fact that $p$ contributes
as an additional part of the hopping term.  Within the MF treatment, the value
of $\tilde{J}/\tilde{U}$ for the $g=1$ MI lobe tip is decreased by $8\%$
when $p=0.04$ (a realistic value for experiments -- see
Appendix~\ref{App:continuumlimit}), and by $18\%$ when $p=0.1$.

\subsection{Gutzwiller \emph{ansatz}}
\label{sect:Gutzwiller}

The Gutzwiller variational ansatz~\cite{Rokhsar1991PRB, Krauth1992PRB,
Jaksch1998PRL} is a standard tool to treat bosonic lattice systems, giving
access to both static and dynamic properties of the BH model.  In the
homogeneous version that we employ in this work, the ansatz for the wave
function reads
\begin{equation}\label{17}
\left| G \right\rangle =
\prod_i \sum_{n=0}^{n_\mathrm{max}} f_n \left| n \right\rangle_i,
\end{equation}
where the product runs over all lattice sites.  The coefficients $f_n$ of the
on-site Fock states $\left| n \right\rangle_i$ do not depend on the lattice site
$i$, and $n_\mathrm{max}$ is a cut-off on the maximum on-site occupation number.
Writing the ground state of $\tilde{K}_B$ as a product of single-site states is
only valid as an approximation.  For the ordinary Bose-Hubbard model (with $V =
P = 0$), this approximation becomes exact both in the $\tilde{U} \to 0$ and
$\tilde{J} \to 0$ limits~\cite{Bloch2008RMP} (in the infinite-volume limit and
with $n_\mathrm{max} \rightarrow \infty$).

We use $\left| G \right\rangle$ as a variational \emph{ansatz} for the ground
state of $\tilde{K}_B$, which depends on the $n_\mathrm{max}+1$ variational
parameters $\lbrace f_n \rbrace$.  To extract observables from this
ansatz, one should first find the set of coefficients $\lbrace f_n
\rbrace$ that minimizes the expectation value of the energy per site,
$\mathcal{E}\left( f_0, \dots, f_{n_\mathrm{max}} \right)$.  We perform the
optimization numerically through Simulated Annealing~\cite{Kirkpatrick1983},
which is a stochastic minimization algorithm (an implementation is made publicly
available ~\cite{Comparin2017Zenodo}).

For the ordinary BH model, the optimization of the Gutzwiller \emph{ansatz} is
strictly equivalent to the MF theory based on the
decoupling~\cite{Sheshadri1993EPL}
\begin{equation}
\label{18}
\hat{b}_i^\dagger \hat{b}_j \simeq
\varphi^\ast \hat{b}_j + \hat{b}_i^\dagger \varphi - |\varphi|^2,
\end{equation}
where $\varphi \equiv \langle \hat{b}_i \rangle$ is the condensate order
parameter, assumed to be independent on the site $i$.  The choice of the
decoupling is less straightforward when the Hamiltonian includes additional
terms, as we mention in Section~\ref{sect:boundaries:MF}.  The Gutzwiller variational
procedure, however, remains well defined, so that we use it as the basis of our
analysis.  Once the optimal set of coefficients $\lbrace f_n \rbrace$ is found,
several observables can be computed, and especially the density $\langle
\hat{n}_i \rangle$ and the condensate density $|\varphi|^2$.  As an example (for
a specific choice of $V$ and $P$), the density as a function of
$\tilde{J}/\tilde{U}$ and $\tilde{\mu}/\tilde{U}$ is shown in
Fig.~\ref{fig:phasediags}(c), where we identify the Mott lobes with integer
density. These regions corresponds to those where the condensate density vanishes
(not shown).

For $q > 0$, the boundary between two subsequent Mott lobes is defined on a
finite $\tilde{J}$ interval at $\tilde{\mu} = \tilde{\mu}^{g,g+1}$, extending up
to a triple point [where $\tilde{J} / \tilde{U}$ equals $(\tilde{J} /
\tilde{U})_\mathrm{triple}^{g,g+1}$], and it is characterized by a discontinuous
jump in the density.  Three transition lines cross at the triple point:
The MI/SF transition for the $g$th Mott lobe, the MI/SF transition for the $(g
+ 1)$th Mott lobe, and the MI/MI transition between the $g$th and $(g + 1)$th
Mott lobes (see Fig.~\ref{fig:phasediags})

To locate the triple point, we simplify the Gutzwiller ansatz, so that it
can be treated analytically.  Close to the transition between two subsequent
Mott phases with filling $g$ and $g+1$, it is a valid approximation to only
consider the two on-site Fock states $|g\rangle$ and $|g+1\rangle$, which
results in the \emph{ansatz}
\begin{equation}\label{21}
\left| G_2 \right\rangle =
\prod_i \left[
\cos\theta |g\rangle_i +
\sin\theta |g + 1\rangle_i
\right],
\end{equation}
where $\theta$ is the only variational parameter.  For $\theta=0$ and
$\theta=\pi/2$, this state corresponds to a product of local Fock states with
filling equal to $g$ and $g + 1$, respectively. This corresponds to a MI state.
When $0 < \theta < \pi / 2$, on the contrary, each local state is mixed, and the
average value of $\hat{b}_i$,
\begin{equation}
\label{22}
\langle \hat{b}_i \rangle \equiv \cos\theta \sin\theta \sqrt{g+1} ,
\end{equation}
takes a non-zero value, which corresponds to a SF state [within the
approximation in Eq.~\eqref{21}].
As in the more general case of $|G\rangle$, the variational approach requires
finding the value of $\theta$ which minimizes the average energy per site
$\mathcal{E}(\theta)$.
At the MI/MI transition, $\mathcal{E}(\theta)$ shows the double-well shape that is
peculiar of discontinuous transitions, where an energy barrier separates two
local minima with the same energy. If $\tilde{\mu}$ is slightly lower (higher)
than $\tilde{\mu}^{g,g+1}$, the minimum in $\theta = 0$ ($\theta = \pi / 2$)
becomes the global minimum, and the ground state is a Mott insulator with $g$
($g + 1$) atoms per site.

When the chemical potential is tuned to its boundary value ($\tilde{\mu} =
\tilde{\mu}^{g,g+1}$), the energy profile is symmetric: $\mathcal{E}(\theta) =
\mathcal{E}(\pi/2 - \theta)$, and the height of the energy barrier between the
two minima equals $\Delta\mathcal{E} = \mathcal{E}(\pi/4) - \mathcal{E}(0)$.
This barrier height decreases when $\tilde{J} / \tilde{U}$ increases, and the
triple point is reached when $\Delta \mathcal{E} = 0$. This condition can be
explicitly rewritten as
\begin{equation}
\label{23}
\left( \frac {z \tilde{J}} {\tilde{U}} \right)_\mathrm{triple}^{g,g+1}
= \frac{q}{2 (1 + g) (1 + p + 2 p g)}.
\end{equation}
When $\tilde{J} / \tilde{U}$ increases beyond the triple point, the minimum of
$\mathcal{E}$ takes place at a finite value of the variational parameter
$\theta$, corresponding to the SF phase.  The triple point in Eq.~\eqref{23},
computed within the two-states Gutzwiller \emph{ansatz}, agrees with the values
found via the numerical data obtained with the full Gutzwiller scheme; see
Fig.~\ref{fig:phasediags}(c).

The MI/SF phase transition is characterized by number fluctuations and the
two-states approximation $|G_2\rangle$ is not
sufficient to identify the corresponding boundaries away from the triple point.
To this purpose, we consider the three-states \emph{ansatz}
\begin{align}\label{Gutz3}
\left| G_3 \right\rangle =
\prod_i & \left[
\cos\theta_1 \sin\theta_2 | g-1 \rangle_i +
\cos\theta_1 \cos\theta_2 | g \rangle_i \right
.
\nonumber\\
&\left.
+ \sin\theta_1 | g+1 \rangle_i
\right],
\end{align}
where $\theta_1$ and $\theta_2$ are variational parameters. In the $g$th Mott lobe
one has $\theta_1 = \theta_2 = 0$. To the lowest order in powers of
$\theta_1$
and $\theta_2$, the energy per site corresponding to the ansatz \eqref{Gutz3} is
\begin{equation}
\mathcal{E}(\theta_1, \theta_2) = \mathcal{E}_0 + \sum_{a,b=1,2} \theta_a M_{ab}
\theta_b\,,
\end{equation}
where
\begin{align}\label{Eq:eo}
\mathcal{E}_0 &= g(g-1)\frac{\tilde U}{2}-\frac{g^2 zV}{2}-g\tilde\mu \,,\nonumber\\
M_{11} & = g\tilde U - (g+1)z\tilde J- (2g^2+3g+1)zP -gzV-\tilde\mu  \,, \nonumber\\
M_{22} & = (1-g)\tilde U - g z\tilde J- g(2g-1)zP + gzV+\tilde\mu \,,\nonumber\\
M_{12} &= M_{21} = - \sqrt{g(g+1)}z\tilde J - 2g\sqrt{g(g+1)} zP\,.
\end{align}
The phase boundaries are then obtained through the saddle point condition, namely by
solving $\textrm{det}\,M=0$ (out of the two solutions, the one with
smallest $\tilde{J}$ should be considered). The resulting analytical expression
for the transition lines matches with the numerical phase diagram computed
through the Gutzwiller \emph{ansatz}.  In particular, the tip of the lobe is
located at the critical value
\begin{equation}
\left(\frac{z\tilde{J}}{\tilde U}\right)_c =
\frac{1}{p + (1 + 2 g p) \left(2 g + 1 + 2 \sqrt{g (g + 1)}\right)} .
\label{eq:lobe_tip_gutzwiller}
\end{equation}
This value depends on $p$, and the effect is larger for MI regions with larger
filling $g$~\cite{Dutta2015RPP}.
We stress that the nearest-neighbor attraction $q$, on the contrary, only
affects the position of the lobe tip along the chemical-potential axis, and not
the value of $(z\tilde{J}/\tilde{U})_c$.

\subsection{Mean-field theory}
\label{sect:boundaries:MF}

Following the conventional approach used for the BH model, we also compute the
phase diagram through a perturbative MF analysis \cite{Fisher1989PRB,
Jaksch1998PRL, VanOosten2001PRA} (an alternative approach is described in
Appendix~\ref{App:lobes}).
Similarly to Eq.~\eqref{18}, we decouple the density-dependent hopping term as
\begin{equation}
\hat{b}_i^\dagger \hat{n}_i \hat{b}_j
\simeq
\varphi \hat{b}_i^\dagger \hat{n}_i + \varphi \hat{n}_i \hat{b}_j -
\varphi^2 \hat{n}_i,
\end{equation}
assuming that $\varphi$ is real.  This allows us to write the Hamiltonian in
Eq.~\eqref{11} as $\tilde{K}_B = \tilde{K}_B^{(0)} + \varphi \tilde{K}_B^{(1)}$,
with
\begin{align}
\tilde{K}^{(0)}_{B}& =
\frac{\tilde{U}}{2} \sum_{i}\hat{n}_{i} (\hat{n}_{i}-1)
-
\tilde{\mu} \sum_{i} \hat{n}_{i}
-
\frac{V}{2} \sum_{\langle i,j\rangle} \hat{n}_{i} \hat{n}_{j}
\nonumber\\
&\quad +
z\tilde{J} \varphi^{2} \left[
N_{s} (1 + p) + 2 p \sum_{i} \hat{n}_{i} \right],
\end{align}
and
\begin{align}
\tilde{K}^{(1)}_{B}&
=
-z\tilde{J}\sum_{i}
\left\lbrace
[1+p(\hat{n}_{i}+1)]
\hat{b}_{i}+\hat{b}^{\dag}_{i}[1+p(\hat{n}_{i}+1)]\right\rbrace
\nonumber \\
&\quad -p\tilde{J}
\sum_{\langle i,j\rangle}(\hat{n}_i \hat{b}_j + \hat{b}^{\dag}_{i} \hat{n}_{j}),
\end{align}
with $N_s$ being the number of sites.
$\tilde{K}_B^{(0)}$ is diagonal in the Fock basis, while $\tilde{K}_B^{(1)}$
only includes off-diagonal terms.
Due to the presence of nearest-neighbors interactions and
density-dependent hopping, neither $\tilde{K}_B^{(0)}$ nor $\tilde{K}_B^{(1)}$
are defined on a single site, at a difference with the ordinary MF theory for
the BH model.
To identify the MI/SF transition, we study the stability of the homogeneous Fock
state $|\Phi_g \rangle = |g,\dots,g\rangle$, by treating $\varphi
\tilde{K}_B^{(1)}$ as a perturbative correction. For $\tilde{J}=0$, $\varphi$
vanishes and the unperturbed ground state is $|\Phi_g\rangle$, for
$\tilde{\mu}^{g-1,g} < \tilde{\mu} < \tilde{\mu}^{g,g+1}$
\cite{Chaviguri2017PRA}.
We compute the total energy at second order in perturbation theory
\cite{VanOosten2001PRA}, $E(g) = E_{0}(g) + \varphi^{2} E_{2}(g)$, as
\begin{equation}
E_{2}(g) =
\sum_{k}
\sum_{\alpha=P,H}
\frac
{|\langle\Phi_g | \tilde{K}_B^{(1)} | \Phi^{k,(\alpha)}_g \rangle|^2}
{E_0(g) - E^{(\alpha)}_0(g)},
\end{equation}
considering the following excited states with an additional particle (P) or hole
(H) on site $k$: $|\Phi^{k,(P)}_g \rangle = \frac {\hat{b}^\dag_k} {\sqrt{g +
1}} | \Phi_g \rangle$ and $|\Phi^{k,(H)}_g \rangle = \frac {\hat{b}_k}
{\sqrt{g}} | \Phi_g \rangle$.  The energy of the unperturbed state reads
$E_{0}(g)=N_{s} \mathcal{E}_{0} +z\tilde{J}\varphi^{2}N_{s}(1+p+2pg)$ [see
Eq.~\eqref{Eq:eo}].

The MI lobe boundary corresponds to the condition that the coefficient of
$\varphi^2$ in $E(g)$ vanishes, which yields
\begin{equation}
\label{eq:lobes_MF_T0}
\begin{aligned}
\frac{z\tilde{J}}{\tilde{U}} &= (1+p(2g+1)) \\
&\times\left[
\frac{(g+1)(1+p(2g+1))^2}{\Delta\mathcal{E}_P} +
\frac{g(1+2pg)^2}{\Delta\mathcal{E}_H}
\right]^{-1},
\end{aligned}
\end{equation}
where the energy gaps
\begin{align}
\Delta \mathcal{E}_P &= \tilde{U} g - \tilde{\mu} - z V g,\\
\Delta \mathcal{E}_H &= -\tilde{U} (g-1) + \tilde{\mu} + z V g,
\end{align}
are associated to single-site particle/hole excitations at $\tilde{J}=0$.

By comparing the MF analytical phase diagram with the one obtained through the
$|G_3\rangle$ Gutzwiller ansatz, we find that the two approaches are essentially
equivalent.  As an example, we consider the value of $\tilde{J}/\tilde{U}$ at
the tip of the $g=1$ lobe [extracted from Eq.~\eqref{eq:lobes_MF_T0}] and
compare it with the Gutzwiller result in Eq.~\eqref{eq:lobe_tip_gutzwiller}: For
$p$ as large as $0.3$, the relative deviation is as small as $0.5\%$, and the
difference becomes even smaller when lobes with larger $g$ are considered.

\section{Conclusions}
\label{sect:conclusions}

We determined the quantum phase diagram of ultracold bosonic impurities immersed
in a vortex lattice.  The effective EBH model (derived via the polaron
transformation, with parameters chosen through the variational method), takes
into account the lattice excitations represented by Tkachenko modes. These
excitations generate peculiar terms in the resulting EBH model, namely a
long-range attractive potential and a density-dependent hopping.  The
corresponding coefficients $V$ and $P$ are enhanced, as compared to ordinary
optical-lattice realizations of the Bose-Hubbard model.  The phase diagram
includes Mott-insulator and superfluid phases, and it is computed with two
independent techniques (a Gutzwiller variational approach and a perturbative
mean-field theory). The separate effects of $V$ and $P$ on the MI lobes are
clearly identified: A nonvanishing $V$ leads to the appearance of
discontinuous MI/MI transitions and of a triple point, while a non-zero $P$
induces a shift of the critical hopping parameter for the SF/MI transition.  The
triple point can also be recognized in the exact phase diagram of the model for
$P=0$, previously computed through the quantum Monte Carlo technique
\cite{Chaviguri2017PRA}.

For the experimental realization of our scheme, we propose to employ the
technique of artificial magnetic fields \cite{Spielman2009PRA}, for which the
generation of vortices is currently being optimized \cite{Price2016NJP} and
could potentially reach the vortex-lattice regime predicted by theory
\cite{Spielman2009PRA, Bai2017Arxiv}.  Finally, although we focused here on
neutral bosons, the atomic trapped species could also consist of fermions with
spin charges, which would allow one to study magnetism with ultracold atoms,
including triangular-lattice frustrated spins.

\begin{acknowledgments}
We acknowledge Chiara Menotti and Elia Macaluso for useful discussions.  This work is supported by
CAPES/PROEX, FAPESP/CEPID, the EU-FET Proactive grant AQuS, Project No. 640800, ERC
Starting Grant TopoCold, and by Provincia Autonoma di Trento.
\end{acknowledgments}

\bibliography{refs}


\onecolumngrid
\appendix

\section{Effective EBH model}
\label{App:EBHmodel}

The EBH in Eq.~\eqref{11} results from the unitary transformation in the
Hamiltonian of Eq.~\eqref{8}, followed by its average over the lattice modes,
i.e., $\tilde{K}_{B} = \langle\tilde{K}_{B}\rangle_\mathrm{ph}$, with
$|\mathrm{ph}\rangle = \prod_{\mathbf{q}} |N_{\mathbf{q}}\rangle$ and
$|N_{\mathbf{q}}\rangle$ being the number state of the lattice modes
\cite[Sections 4.31 and 4.32]{Mahan2000}.  From that, we obtain the
energy coefficients
\begin{align}
\tilde{U}  &=
U+\frac{2}{S}\sum_{\textbf{q}}\Big[|\alpha_{i,\textbf{q}}|^{2}\epsilon_{\textbf{q}}-g_{AB}\sqrt{n_{A}}
\Big(\Omega^{ii}_{\textbf{q}}e^{i\textbf{q}\cdot\mathbf{R}_{i}}\alpha^{\ast}_{i,\textbf{q}}+\Omega^{ii\ast}_{\textbf{q}}
e^{-i\textbf{q}\cdot\mathbf{R}_{i}}\alpha_{i,\textbf{q}}\Big)\Big], \nonumber\\
\tilde{\mu}   &=
\mu_{B}-\frac{1}{S}\sum_{\textbf{q}}\Big[|\alpha_{i,\textbf{q}}|^{2}\epsilon_{\textbf{q}}-g_{AB}\sqrt{n_{A}}
\Big(\Omega^{ii}_{\textbf{q}}e^{i\textbf{q}\cdot\mathbf{R}_{i}}\alpha^{\ast}_{i,\textbf{q}}+\Omega^{ii\ast}_{\textbf{q}}
e^{-i\textbf{q}\cdot\mathbf{R}_{i}}\alpha_{i,\textbf{q}}\Big)\Big], \nonumber\\
V &=
-\frac{2}{S}\sum_{\textbf{q}}\Big[\alpha^{\ast}_{j,\textbf{q}}\alpha_{i,\textbf{q}}
e^{-i\textbf{q}\cdot\textbf{d}}\epsilon_{\textbf{q}}-g_{AB}\sqrt{n_{A}}
\Big(\Omega^{jj}_{\textbf{q}}e^{i\textbf{q}\cdot\mathbf{R}_{i}}\alpha^{\ast}_{i,\textbf{q}}
+\Omega^{jj\ast}_{\textbf{q}}e^{-i\textbf{q}\cdot\textbf{R}_{i}}\alpha_{i,\textbf{q}}\Big)\Big],\label{a1}
\end{align}
\begin{align}
\tilde{J}  &= Jf^{0},\nonumber \\
P   &=
\frac{g_{AB}\sqrt{n_{A}}}{S}\sum_{\textbf{q}}\Big[\Omega^{ij}_{\textbf{q}}e^{i\textbf{q}\cdot\textbf{R}_{i}}\alpha^{\ast}_{i,\textbf{q}}+
\bar{\Omega}^{ij}_{\textbf{q}}e^{-i\textbf{q}\cdot\textbf{R}_{i}}\alpha_{i,\textbf{q}}\Big]f^{0},\label{a2}
\end{align}
which depend explicitly on the parameter $\alpha$ of the unitary transformation,
including
\begin{equation}
f^{0}=\exp\bigg[-\frac{1}{2S}
\sum_{\textbf{q}}|\alpha_{i\textbf{q}}|^{2}|1-e^{-i\textbf{q}\cdot\textbf{d}}|^{2}\bigg].
\end{equation}
Using the parameter $\alpha$ obtained through the variational method [see
Appendix~\ref{App:variationalmethod}],  the  energies above are reduced to
Eqs.~(\ref{12}-\ref{16}) in the main text.

\section{Variational method}
\label{App:variationalmethod}


Applying the extremization
condition to the total energy of the system, $E= \langle \tilde{K}_{B}
\rangle_\Phi$, with $|\Phi_{g}\rangle=|g,\ldots,g\rangle$, we get
\begin{equation}\label{b1}
  \frac{\partial E}{\partial\alpha_{i,\mathbf{q}}} \quad \text{implying} \quad
  \frac{\partial\mathcal{E}_{0}}{\partial\alpha_{i,\mathbf{q}}}=0.
\end{equation}

The second condition is justified by using the fact
that $P\thicksim\tilde{J}$ and $V\thicksim\tilde{U}$. Since $U\gg J$, only the
ground
state energy is relevant. Applying Eq.~\eqref{b1} to each element of
the ground-state energy in Eq.~\eqref{Eq:eo}, according to Eq.~\eqref{a1}
\begin{align}
\frac{\partial\tilde{U}}{\partial\alpha_{i,\mathbf{q}}} &=
\frac{2}{S}\Big[\alpha^{\ast}_{i,\mathbf{q}}\epsilon_{\mathbf{q}} -
g_{AB}\sqrt{n_{A}}\Omega^{ii\ast}_{\mathbf{q}}e^{-i\mathbf{q}\cdot\mathbf{R}_{i}}\Big],
\quad
\frac{\partial\tilde{\mu}}{\partial\alpha_{i,\mathbf{q}}} =
-\frac{2}{S}\Big[\alpha^{\ast}_{i\mathbf{q}}\epsilon_{\mathbf{q}} -
g_{AB}\sqrt{n_{A}}\Omega^{ii\ast}_{\mathbf{q}}e^{-i\mathbf{q}\cdot\mathbf{R}_{i}}\Big],
\nonumber\\
\frac{\partial V}{\partial\alpha_{i,\mathbf{q}}} &=
-\frac{2}{S}\Big[\alpha^{\ast}_{j,\mathbf{q}}\epsilon_{\mathbf{q}}e^{-i\mathbf{q}\cdot\mathbf{d}} -
g_{AB}\sqrt{n_{A}}\Omega^{jj\ast}_{\mathbf{q}}e^{-i\mathbf{q}\cdot\mathbf{R}_{i}}\Big].
\end{align}
Assuming homogeneity ($\alpha_{i\mathbf{q}}\equiv\alpha_{j\mathbf{q}}$), we find
\begin{equation}\label{b3}
\alpha_{i,\mathbf{q}}=\frac{g_{AB}\sqrt{n_{A}}}{\epsilon_{\mathbf{q}}}\Omega^{ii}_{\mathbf{q}}e^{i\mathbf{q}\cdot\mathbf{R}_{i}}.
\end{equation}
which depends on the Tkachenko-mode parameters $\Omega^{ii}_{\mathbf{q}}$ and
$\epsilon_\mathbf{q}$.

\section{Density-dependent hopping}
\label{App:densitydependenthopping}

In Ref.~\cite{Luhmann2012NJP}, the authors determined a generalized BH model
by considering the nearest-neighbors contribution in the interacting term.
For ultracold atoms trapped in a three-dimensional
optical lattice, this reads
\begin{equation}
H =- J \sum_{\langle i,j \rangle} \hat{b}_{i}^{\dag} \hat{b}_{j}+ \frac{1}{2}
\sum_{ijkl}
U_{ijkl}\hat{b}^{\dag}_{i}\hat{b}^{\dag}_{j}\hat{b}_{k}\hat{b}_{l}-\mu \sum_{i}
\hat{n}_i,
\end{equation}
whit the total hopping
\begin{equation}
J_{\text{Total}}=\int
d^{3}r\omega^{\ast}(\mathbf{r}_{i})\Big[-\frac{\hbar^{2}\nabla^{2}}{2m}+V(\mathbf{r})+
g\rho(\mathbf{r})\Big]\omega^{\ast}(\mathbf{r}_{j}),
\quad \text{with} \quad
\rho=n_{i}|\omega(\mathbf{r}_{i})|^{2}+(n_{j}-1)|\omega(\mathbf{r}_{j})|^{2},
\end{equation}
and the long-range potential
\begin{equation}
 V=g\int d^{3}r|\omega(\mathbf{r}_{i})|^{2}|\omega(\mathbf{r}_{j})|^{2},
\end{equation}
where $g=4\pi\hbar a_{s}/m$ being the repulsive contact potential and $a_{s}$
the scattering length.  We consider the hopping correction due to the
interaction (in our two-dimensional case $g\equiv g_{B}$) $J_{E}=g_{B}\int
d^{2}r\omega^{\ast}(\mathbf{r}_{i})\rho(\mathbf{r})\omega^{\ast}(\mathbf{r}_{j})$.  By applying the
Gaussian ansatz for the Wannier functions,
$\omega_{B}(\mathbf{r})=|B_{0}|e^{-r^{2}/2l^{2}_{0}}$, with width $l^{2}_{0}=\hbar
\xi_A / \sqrt{m_{B}V_{0}}$ (which corresponds to the harmonic ansatz for
the vortex-core density profile \cite[Chapter 9]{Pethick2008}), we derive
\begin{align}
J_{E}=&g_{B}(n_{i}+n_{j}-1)|B_{0}|^{4}\bigg[\int^{\infty}_{-\infty}dx
\exp[-(x-x_{i})^{2}/2l^{2}_{0}-(x-x_{j})^{2}/2l^{2}_{0}-(x-x_{i})^{2}/l^{2}_{0}]\bigg]^{2}\nonumber\\
=&(n_{i}+n_{j}-1)\frac{g_{B}}{2\pi l^{2}_{0}}e^{-3d^{2}/4l^{2}_{0}}
=(n_{i}+n_{j}-1)Ue^{-6\sqrt{2}\Gamma_{L}\sqrt{V_{0}/E_{R}}},
\end{align}
and
\begin{align}
V_{E}=&g_{B}|B_{0}|^{4}\bigg[\int^{\infty}_{-\infty}dx
\exp[-(x-x_{i})^{2}/l^{2}_{0}-(x-x_{j})^{2}/l^{2}_{0}]\bigg]^{2}
=\frac{g_{B}}{2\pi l^{2}_{0}}e^{-d^{2}/l^{2}_{0}}\nonumber\\
=&Ue^{-8\sqrt{2}\Gamma_{L}\sqrt{V_{0}/E_{R}}},
\end{align}
where $d=x_{i}-x_{j}$, $U_{iiij}=J_{E}$ and $U_{ijij}=V_{E}$.

\begin{figure}[htb]
\centering
\includegraphics[width=0.4\linewidth]{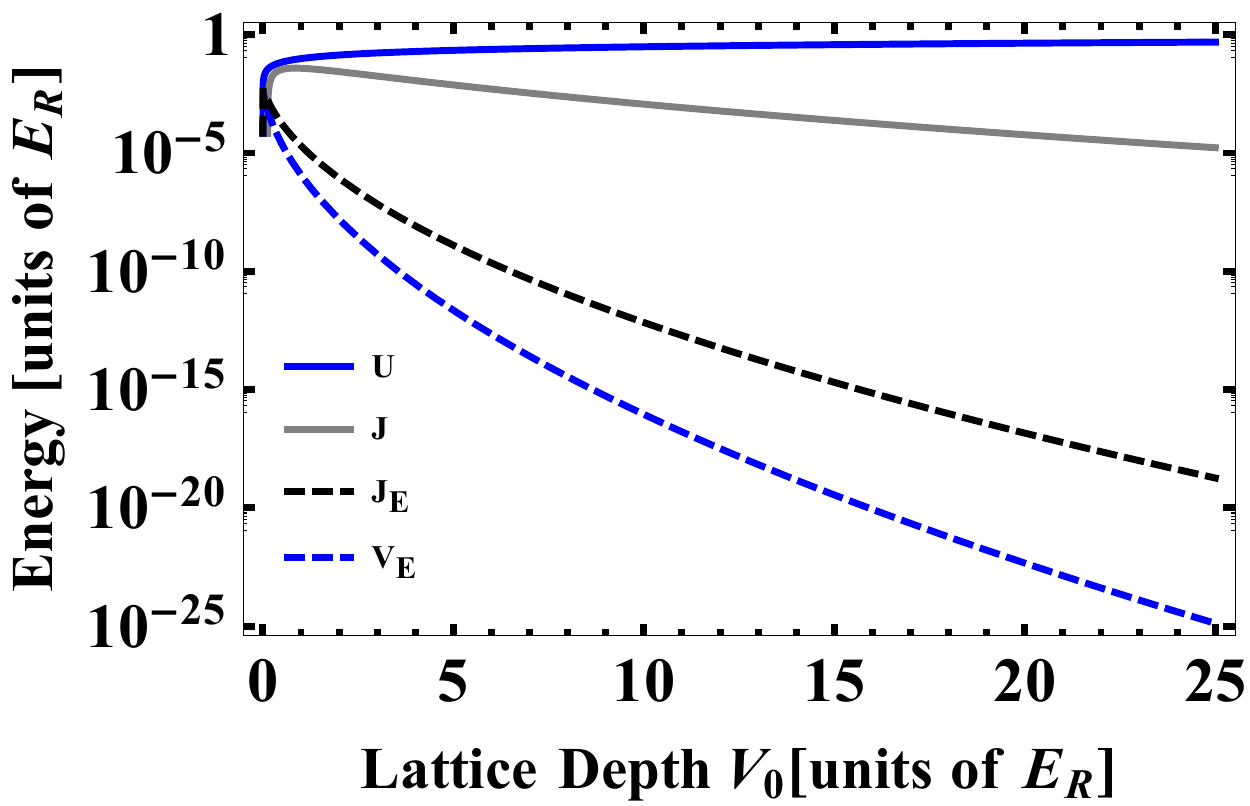}
\caption{From top to bottom: Solid blue: on-site energy $U$. Solid
gray: ``Gaussian'' hopping $J$. Dashed black: induced hopping $J_{E}$.
Dashed blue: long-range potential $V_{E}$.
We set $a_{B}/l^{B}_{z}\approx0.05$
with $\Gamma_{L}\approx1$ for $^{87}$Rb atoms and assume an unitary filling of
the lattice sites, $n_{i}=n_{j}=1$.
\label{f3}}
\end{figure}

In addition, we determine the on-site energy,
$U=E_{R}(2/\sqrt{\pi})(a_{B}/l^{B}_{z})\sqrt{V_{0}/E_{R}}$, and the hopping,
$J =
E_{R}\big[(2\Gamma_{L}-1)(V_{0}/E_{R})-(\sqrt{2}/4)\sqrt{V_{0}/E_{R}}\big]e^{-2\sqrt{2}\Gamma_{L}\sqrt{V_{0}/E_{R}}}$.
By comparing the magnitude of the different terms in Fig.~\ref{f3}, it is clear
that the off-site energies decay faster than $U$ and $J$.  This analysis
justifies the validity of the BH truncation applied in Eq.~\eqref{4}.

On the other hand, by applying the same procedure for the effective EBH in
Eq.~\eqref{11}, where we  associate the hopping
\begin{equation}
\hat{\tilde{J}}_{\text{Total}}=\hat{\tilde{J}}+\hat{J}_{P},
\end{equation}
with $\hat{\tilde{J}}=-\tilde{J}\hat{b}^{\dag}_{i}\hat{b}_{j}$,
$\hat{J}_{P}=-P(\hat{n}_{j}+\hat{n}_{i})\hat{b}^{\dag}_{i}\hat{b}_{j}$ and
$\tilde{J}=Jf^{0}$, we find
\begin{equation}
\hat{\tilde{J}}_{\text{Total}}=-[Jf^{0}+P(\hat{n}_{j}+\hat{n}_{i})]\hat{b}^{\dag}_{i}\hat{b}_{j}.
\end{equation}

Finally, using the ``bare'' hopping from Eq.~\eqref{5} and the explicit form of
the density-dependent hopping $P$ given by Eq.~\eqref{16}, we obtain
\begin{equation}\label{c1}
\hat{\tilde{J}}_{\text{Total}}=f^{0}\int
d^{2}r\omega^{\ast}(\mathbf{r}_{i})\bigg[-\frac{\hbar^{2}\nabla^{2}}{2m_{B}}+\hat{V}_{\text{eff}}\bigg]\omega(\mathbf{r}_{j})
\hat{b}^{\dag}_{i}\hat{b}_{j},
\end{equation}
with $\hat{V}_{\text{eff}}=V_{AB}(\mathbf{r})-V_{0}\hat{\rho}(\mathbf{r})$ and
$\hat{\rho}(\mathbf{r})=\mathcal{X}(\mathbf{r})(\hat{n}_{j}+\hat{n}_{i})$ being the effective
potential associated with density-dependent hopping and reduced density,
respectively.  The parameter $\mathcal{X}(\mathbf{r}) =
(g_{AB}/S)\sum_{\mathbf{q}}[(\varphi_{A} u^{\ast}_{\mathbf{q}} -
\varphi^{\ast}_{A} v_{\mathbf{q}}) \Omega^{ii}_{\mathbf{q}} + c.c.]
(1/\epsilon_{\mathbf{q}})$ represents the impurity spatial density ``dressed''
by the lattice modes, with $\Omega^{ii}_{\mathbf{q}}$ given by Eq.~\eqref{9}.

\section{EBH parameters in the continuum limit}
\label{App:continuumlimit}

The dispersion relation of the Tkachenko modes in the continuum limit is
established by considering small values of momentum, that is, $ql\ll1 $
\cite{Caracanhas2013PRL}.
 In addition, we consider
the energy dispersion of the Tkachenko modes $\epsilon_{q} \approx
\hbar^2q^2/2M$, with $M= \frac{1}{2\kappa\sqrt{\eta}}\frac{\hbar\Omega}{n_A
g_A}m_A$ being the effective mass of the modes  and the lattice constants
$\kappa = 1.1592$ and  $\eta = 0.8219$. In the
low-energy limit, we have $u_{\mathbf q}(\mathbf r)\approx \varphi_A(\mathbf
r)\; c_{1\mathbf q}\,e^{i\mathbf q\cdot \mathbf r}$ and $v_{\mathbf q}(\mathbf
r)\approx \varphi_A(\mathbf r)\; c_{2 \mathbf q}\,e^{-i\mathbf q\cdot \mathbf
r}$. The small value of the momentum allows us to expand $(c_{1q}-c_{2q})
\approx
\frac{1}{\sqrt{2}}\,\eta^{1/4}\,(ql)$. Using these considerations and the
Gaussian ansatz for the Wannier functions (see Appendix~\ref{App:densitydependenthopping}), we
determine the $\Omega$ and $\bar{\Omega}$ in Eq.~\eqref{9}, which leads to
\begin{align}
|\Omega^{ii}_{\mathbf{q}}|^{2}&=\frac{\eta^{1/2}}{2}(ql)^{2}\Big[\frac{|B_{0}|}{2
\xi_A^{2}}\int
d\mathbf{r}e^{\pm i\mathbf{q}\cdot\mathbf{r}}
e^{-r^{2}/2l^{2}_{0}}\Big]^{2}\nonumber\\
&=\frac{\eta^{1/2}}{128}(ql)^{2}\bigg(\frac{l^{2}_{0}}{\xi_A^{2}}\bigg)^{2}\Big(4-q^{2}l^{2}_{0}\Big)^{2}e^{-q^{2}l^{2}_{0}/2}\label{d1},
\end{align}
where the sign $\pm$ is associated with $\Omega^{ii}_{\mathbf{q}}$ and
$\bar{\Omega}^{ii}_{\mathbf{q}}$ respectively. In the same way, we obtain
\begin{align}
\Omega^{ij}_{\mathbf{q}}&=\frac{\eta^{1/4}}{\sqrt{2}}(ql)\Big[|B_{0}|\int
d\mathbf{r}e^{\pm i\mathbf{q}\cdot\mathbf{r}}
e^{-r^{2}/2l^{2}_{0}}e^{-|\mathbf{r}+\mathbf{d}|^{2}/2l^{2}_{0}}\Big]\nonumber\\
&=\frac{1}{\sqrt{2}}\eta^{1/4}(ql)e^{-d^{2}/4l^{2}_{0}}e^{-q^{2}l^{2}_{0}/4}e^{-i\mathbf{q}\cdot\mathbf{d}/2}\label{d2},
\end{align}
where, we find $\bar{\Omega}^{ij}_{\mathbf{q}}=\Omega^{ij\ast}_{\mathbf{q}}$ for
the continuum case.\\

The long-range potential is obtained by substituting the above relations into the
Eq.~\eqref{15}
\begin{align}
V&=\frac{2g^{2}_{AB}n_{A}}{S}\sum_{\mathbf{q}}\frac{1}{\epsilon_{q}}\Big[\frac{\eta^{1/2}}{128}(ql)^{2}\bigg(\frac{l^{2}_{0}}{\xi_A^{2}}\bigg)^{2}
\Big(4-q^{2}l^{2}_{0}\Big)^{2}e^{-q^{2}l^{2}_{0}/2}e^{-i\mathbf{q}\cdot\mathbf{d}}\Big]\nonumber\\
&\sim \int
d\mathbf{q}(4-q^{2}l^{2}_{0})^{2}e^{-q^{2}l^{2}_{0}/2}e^{-i\mathbf{q}\cdot\mathbf{d}}\sim
Ue^{-d^{2}/2l^{2}_{0}}\nonumber\\
&=U\gamma_{1}\Big(\frac{V_{0}}{E_{R}}\Big)\Big(1+16\Gamma^{2}_{L}\frac{V_{0}}{E_{R}}\Big)e^{-4\sqrt{2}\Gamma_{L}\sqrt{V_{0}/E_{R}}},
\label{d3}
\end{align}
where $\gamma_{1}=(1/4\kappa)(m_{A}/m_{B})^{3/2}(a_{A}/a_{B})$.

The density-dependent hopping is determined by plugging relations \eqref{d1} and \eqref{d2}
into Eq.~\eqref{16}
\begin{align}
P&=\frac{g^{2}_{AB}n_{A}}{S}\sum_{\mathbf{q}}\frac{f^{0}}{\epsilon_{q}}\Big[\frac{\eta^{1/4}}{\sqrt{2}}(ql)^{2}\bigg(\frac{l^{2}_{0}}{8\xi_A^{2}}\bigg)
\Big(4-q^{2}l^{2}_{0}\Big)^{2}e^{-q^{2}l^{2}_{0}/4}e^{-d^{2}/4l^{2}_{0}}\cos(\mathbf{q}\cdot\mathbf{d}/2)\Big]\nonumber\\
&\sim f^{0}e^{-d^{2}/4l^{2}_{0}}\int d\mathbf{q}(4-q^{2}l^{2}_{0})
e^{-q^{2}l^{2}_{0}/2}\cos(\mathbf{q}\cdot\mathbf{d}/2)\sim
f^{0}Ue^{-3d^{2}/8l^{2}_{0}}\nonumber\\
&=\sqrt{2}U\gamma_{1}\Big(\frac{V_{0}}{E_{R}}\Big)^{3/2}\Big(1+\sqrt{2}\Gamma_{L}\sqrt{\frac{V_{0}}{E_{R}}}\Big)
e^{-(3\sqrt{2}+\gamma_{2})\Gamma_{L}\sqrt{V_{0}/E_{R}}},
\label{d4}
\end{align}
where $\gamma_{2} = (1/8 \sqrt{\pi} \kappa^{2} \sqrt{\eta}) (m^{2}_{A} / m_{AB}
m_{B}) (a_{A} / l^{B}_{z})$.  Here we note, in both cases, that $V$ and $P$
decay with the intervortex distance $d$. $f^{0}$ is obtained in the same way as
$V$ and $P$. Following the MF approach  of Ref.~\cite{Luhmann2012NJP}, we derive
\begin{equation}\label{d5}
J_{P}= (n_{i}+n_{j})P.
\end{equation}

\begin{figure}[htb]
\centering
\includegraphics[width=0.4\linewidth]{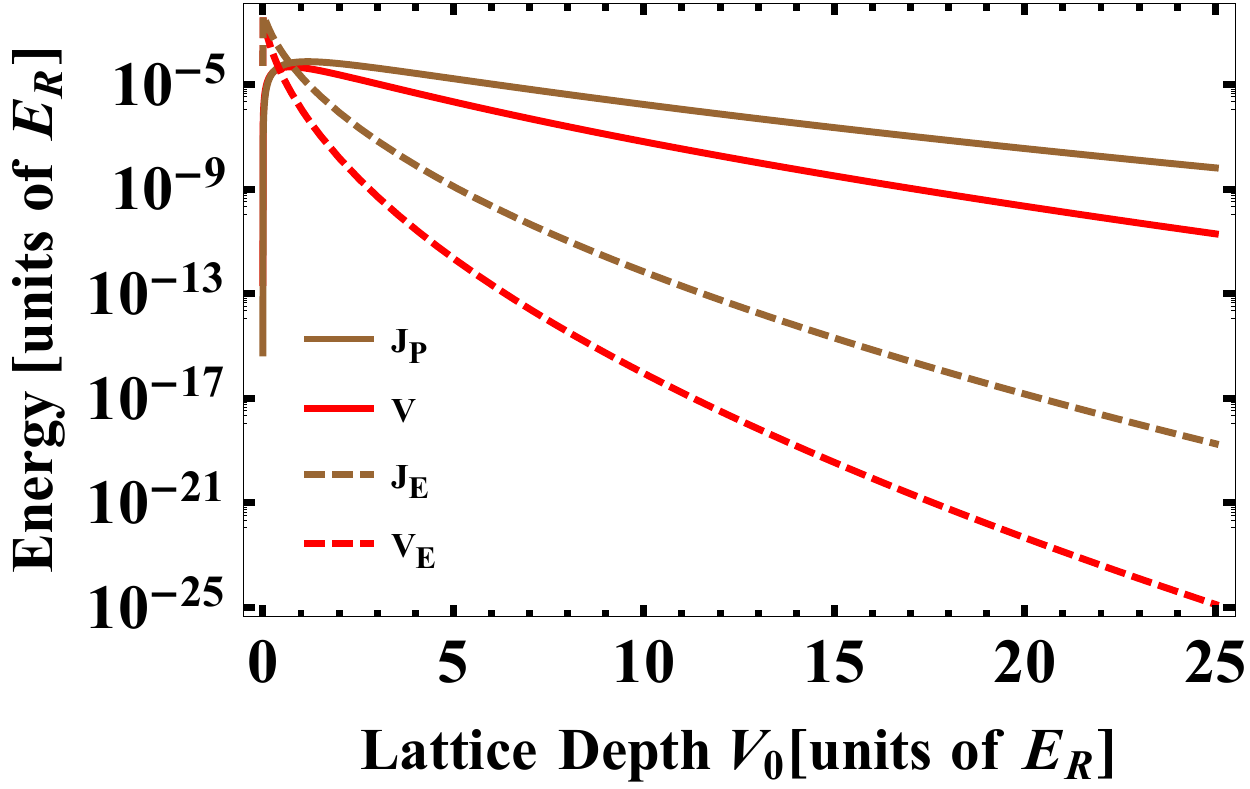}
\caption{From top to bottom: Solid brown: ``dynamic'' hopping $J_{P}$. Solid
red: long-range potential $V$. Dashed brown: ``static'' induced hopping $J_{E}$.
Dashed red: long-range ``static'' potential $V_{E}$.  We use the mixture Na-Rb,
where Rb atoms are considered as impurities. In addition, we assume a unitary
filling of the lattice sites ($n_{i}=n_{j}=1$) and $\Gamma_{L}\approx1$.
\label{f4}}
\end{figure}

In Fig.~\ref{f4}, we compare the long-range potential and the
induced-hopping for the ``static'' and ``dynamic'' case.

In order to determine numerical values of the long-range potential and
density-dependent hopping, we consider the mixture $^{23}$Na  and $^{87}$Rb,
where $^{23}$Na ($m_{A}=23$\emph{u}) and $^{87}$Rb ($m_{B}=87$\emph{u}) are the
majority ($A$) and minority species ($B$), respectively, with $u \simeq 1.66
\times10^{-27}$ kg being the atomic mass unit. For the sodium BEC , we assume an
effective 2D atomic density of $n_A \approx (10^{20} / $m$^3) \times l^A_z$,
with $l^A_z(\sim0.2\mu$m$)$, with $\omega_{z} \sim 2 \pi \times 10 $kHz, being
the axial confinement and where the scattering length of species $A$ is $a_A = 90 a_0$
(with $a_0$ being the Bohr radius).  This yields a unitary Landau factor $n_A
g_A / 2 \hbar \Omega \sim 1$, the lattice parameter $d \sim 0.65 \mu$m, the
vortex size $\xi_A \sim 0.3$ $\mu$m and the Wannier-function length $l_0 \sim
0.2$ $\mu$m. With these parameter values, the tight-binding regime is satisfied:
$V_0/E_R \sim 15 \gg 1$.  Finally, we estimate the values for long-range
potential and density-dependent hopping assuming reasonable values for the
scattering lengths of species $B$, $a_B = 100 a_0$, and for the inter-species
scattering length $a_{AB} = 450 a_0$.  This allows us to be close to the MI/SF
boundary, yielding $zV / \tilde{U} \sim 0.01(z=6)$ and $P / \tilde{J} \sim
0.04$.

\section{Mean-field phase boundaries}
\label{App:lobes}

In this appendix, we derive an alternative expression for the mean-field phase
boundaries computed in Section~\ref{sect:boundaries:MF}, by applying
the method developed for the dipolar EBH model~\cite{Menotti2007PRL}.
On top of $\varphi_i = \langle \hat{b}_i \rangle$, we introduce the average
on-site density $n_i = \langle \hat{n}_i \rangle$.  Analogously to the
decoupling in Eq.~\eqref{18}, we write the other off-site terms in
Eq.~\eqref{11} as
\begin{align}
(\hat{n}_{i}+\hat{n}_{j})\hat{b}^{\dag}_{i}\hat{b}_{j}&\approx \varphi_{j}\hat{b}^{\dag}_{i}\hat{n}_{i}+\varphi^{\ast}_{i}\hat{n}_{j}\hat{b}_{j}
+\hat{b}^{\dag}_{i}\varphi_{j}+\hat{b}_{j}\varphi^{\ast}_{i}-\varphi^{\ast}_{i}\varphi_{j}+\varphi_{j}\hat{b}^{\dag}_{i}\hat{n}_{j}+
\varphi^{\ast}_{i}\hat{n}_{i}\hat{b}_{j}-\varphi^{\ast}_{i}\varphi_{j}(\hat{n}_{i}+\hat{n}_{j}),
\nonumber\\
\hat{n}_{i}\hat{n}_{j}&\approx n_{j}\hat{n}_{i}+n_{i}\hat{n}_{j}-n_{i}n_{j}.
\end{align}

In order to obtain a single-site Hamiltonian, we define
$\varPhi_{i} = \sum_{\langle i\rangle_{j}} \varphi_{j}$,
$N_{i} =\sum_{\langle i\rangle_{j}}n_{j}$ and
$\varPhi_{i}\hat{n}_{i}= \sum_{\langle i\rangle_{j}}\varphi_{j}\hat{n}_{j}$,
where the sums of $\varphi_j$ and $n_j$ over
the nearest neighbors of site $i$.  We also neglect terms which are of second order in $\varphi_i$, the Hamiltonian in
Eq.~\eqref{11} becomes $\tilde{K}_{B} = \tilde{K}^{0}_{B}+\tilde{K}^{1}_{B}$,
where
\begin{align}
\tilde{K}^{0}_{B}& =\sum_{i}\bigg[\frac{\tilde{U}}{2}\hat{n}_{i}(\hat{n}_{i}-1) -
\tilde{\mu}\hat{n}_{i} -\frac{V}{2}(2\hat{n}_{i}-n_{i})N_{i}\bigg],\nonumber\\
\tilde{K}^{1}_{B}& =-\sum_{i}\{[\tilde{J}+P(2\hat{n}_{i}+1)]
\hat{b}_{i} \varPhi^{\ast}_{i} +\hat{b}^{\dag}_{i} \varPhi_{i}
[\tilde{J}+P(2\hat{n}_{i}+1)]\}.
\end{align}

\begin{figure}[htb]
\centering
\includegraphics[width=0.7\linewidth]{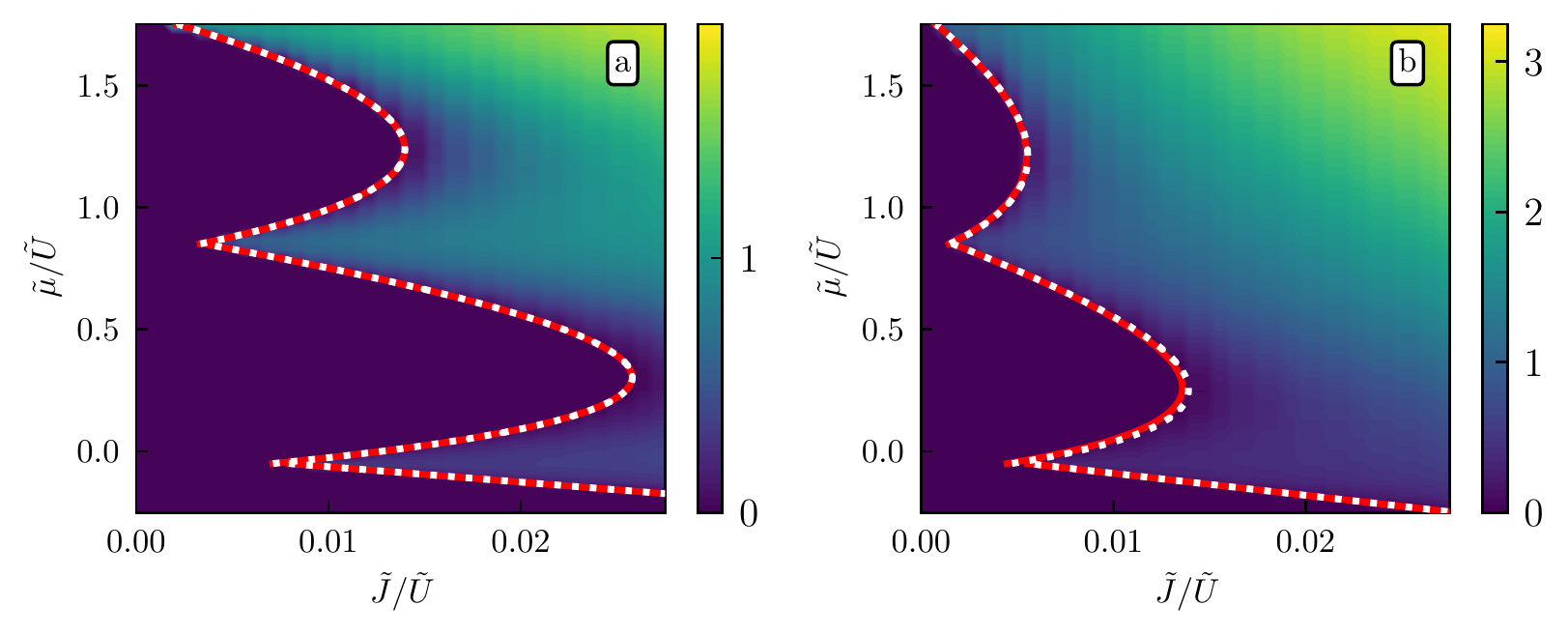}
\caption{
Phase diagram of the EBH model at $q=0.1$, with $p=0.05$ (a) and $p=0.5$ (b).
The condensate density is computed through the Gutzwiller approach (color code,
obtained with $n_\mathrm{max}=6$).
Also shown are the standard MF theory (solid red line,
Eq.~\eqref{eq:lobes_MF_T0}) and the single-site MF approach (dotted white line,
Eq.~\eqref{e5}).
\label{comp}}
\end{figure}

We consider $\tilde{K}^{0}_{B}$ as the main contribution to the ground-state
energy and $\tilde{K}^{1}_{B}$ as a perturbation.  This perturbative MF method
identifies the region where a given density distribution $|\Psi\rangle =
\prod_{i} |n_i \rangle$ is stable against particle and hole excitations, which
corresponds to the Mott lobe.
The method is formulated at finite temperature, followed by the zero temperature limit.
We assume that $|\Psi\rangle$ is a local minimum of $\tilde{K}^{0}_{B}$,
with $\tilde{K}^{0}_{B} |\Psi\rangle = E_{0}|\Psi\rangle$ and
\begin{eqnarray}
\label{e2}
E_{0} = \sum_{i}\mathcal{E}_{0}(n_{i})= \sum_{i} \Big[\frac{\tilde{U}}{2} n_i (n_i - 1)- \tilde{\mu} n_i- \frac{V}{2} n_i N_i\Big].
\end{eqnarray}
At finite inverse temperature $\beta = 1/(k_B T)$ (where $k_B$ is the Boltzmann constant), the order
parameters $\varphi_i$ can be obtained as $\varphi_{i} = \mathrm{Tr} [\hat{b}_i
\hat{\rho}]$.  The density matrix of the system is defined as $\hat{\rho} =
e^{-\beta \tilde{K}_B} / Z$, with
the partition function $Z = \mathrm{Tr} [e^{-\beta \tilde{K}_B}]$, which we approximate by the first term of the Dyson series: $Z \simeq \mathrm{Tr}[e^{-\beta E_{0}}]$.  The Dyson expansion is also applied to calculate
the order parameter, keeping only the lowest-order term
and projecting it in the subspace formed by the states $|\gamma_{1}\rangle =
|\Psi\rangle$ and $|\gamma_{2}\rangle = (\hat{b}_{i}/\sqrt{n_{i}})|\Psi\rangle$.  Then, $\varphi_{i} = \mathrm{Tr} [\hat{b}_{i} e^{-\beta E_0}
e^{-\beta\tilde{K}_B}]$ becomes
\begin{align}
\varphi_{i} &\approx
-e^{\beta E_{0}} \int^{\beta}_{0} d\tau\sum_{|\gamma\rangle}\langle\gamma|
\hat{b}_{i} e^{-(\beta-\tau) \tilde{K}^{0}_{B}} \tilde{K}^{1}_{B} e^{-\tau \tilde{K}^{0}_{B}}|
\gamma \rangle\label{e3}
,
\end{align}
\begin{align}
\varphi_{i}&\approx e^{\beta
E_{0}}\int^{\beta}_{0}d\tau\sum^{|\gamma_{2}\rangle}_{|\gamma\rangle=|\gamma_{1}\rangle}
\langle\gamma|\hat{b}_{i}\sum_{k}e^{-(\beta-\tau)\tilde{K}^{0}_{B}}\Big\{
[\tilde{J}+P(2\hat{n}_{k}+1)]\hat{b}_{k}\varPhi^{\ast}_{k}
+
\hat{b}^{\dag}_{k}[\tilde{J}+P(2\hat{n}_{k}+1)]\varPhi_{k}\Big\}e^{-\tau\tilde{K}^{0}_{B}}|\gamma\rangle
\nonumber\\
&\approx\varPhi_{i}\Big\{(n_{i}+1)[\tilde{J}+P(2n_{i}+1)]\int^{\beta}_{0}d\tau
e^{-(\beta-\tau)(E_{+}-E_{0})}+
n_{i}[\tilde{J}+P(2n_{i}-1)]\int^{\beta}_{0}d\tau
e^{-\tau(E_{-}-E_{0})}\Big\}.
\end{align}
Taking the zero-temperature limit ($\beta\rightarrow\infty$), we derive the equation for the lobe boundaries
\begin{equation}
\varphi_{i}\approx\varPhi_{i}\Bigg[\frac{[\tilde{J}+P(2n_{i}+1)](n_{i}+1)}{E_{+}-E_{0}}+
\frac{[\tilde{J}+P(2n_{i}-1)]n_{i}}{E_{-}-E_{0}}\Bigg].
\end{equation}
The energies are computed from $E_{+}=\langle\gamma_{3}|\tilde{K}^{0}_{B}|\gamma_{3}\rangle$ and
$E_{-}=\langle\gamma_{2}|\tilde{K}^{0}_{B}|\gamma_{2}\rangle$, with
$|\gamma_{3}\rangle=(\hat{b}^{\dag}_{i}/\sqrt{n_{i}+1})|\Psi\rangle$.
Then $E_{+}-E_{0}=\tilde{U}n_{i}-\tilde{\mu}n_{i}-VN_{i}$ and
$E_{-}-E_{0}=-\tilde{U}(n_{i}-1)+\tilde{\mu}n_{i}+VN_{i}$.
Using these relations, we obtain
\begin{equation}\label{e4}
\varphi_{i}\approx\varPhi_{i}\Bigg[\frac{[\tilde{J}+P(2n_{i}+1)](n_{i}+1)}{\tilde{U}n_{i}-\tilde{\mu}-VN_{i}}+
\frac{[\tilde{J}+P(2n_{i}-1)]n_{i}}{-\tilde{U}(n_{i}-1)+\tilde{\mu}+VN_{i}}\Bigg].
\end{equation}
The validity of Eq.~\eqref{e4} is restricted to the interval
\begin{equation}
\tilde{U}(n_{i}-1)-VN_{i}\leq\tilde{\mu}\leq\tilde{U}n_{i}-VN_{i}.
\end{equation}
Considering the approximations $\varPhi_{i}=z\varphi_{i}$ and $N_{i}=zn_{i}$, and
assuming uniformity ($\varphi_{i}=\varphi$ and $n_{i}=g$), we rewrite the Eq.~\eqref{e4}  as
\begin{equation}
\label{e5}
\frac{\tilde{U}}{z\tilde{J}} =
\frac{[1 + p (2 g + 1)] (g + 1)}{g - \tilde{\mu}/\tilde{U} - q g}+
\frac{[1 + p (2 g - 1)] g}      {-(g - 1)+\tilde{\mu}/\tilde{U} + q g}.
\end{equation}
From the expression for the lobe boundary, we can identify the critical point,
that is, the tip of the lobe. This point satisfies
\begin{equation}
\bigg(\frac{z\tilde{J}}{\tilde U}\bigg)_c=
\bigg[
1 + p + 2g (1 + p+ 2pg)+2 \sqrt{g(1+g)(1+p(2g+1))(1+p(2g-1))}\bigg]^{-1}.
\end{equation}
This alternative derivation for the lobe boundary provides results essentially
equivalent to the other two approaches (the Gutzwiller \emph{ansatz} and the
standard MF method).  The comparison is shown in Fig.~\ref{comp}, for both small
and large values of $p$.

\end{document}